**The Transformation of Patient-Clinician Relationships With AI-Based Medical Advice:
A "Bring Your Own Algorithm" Era in Healthcare**

Oded Nov*[a], Yindalon Aphinyanaphongs [a], Yvonne W. Lui [a], Devin Mann [a], Maurizio Porfiri [a], Mark Riedl [b], John-Ross Rizzo [a], Batia Wiesenfeld [a]

*Corresponding Author; [a] New York University; [b] Georgia Institute of Technology

**Introduction**

One of the dramatic trends at the intersection of computing and healthcare has been patients' increased access to medical information, ranging from self-tracked physiological data to genetic data, tests, and scans. Increasingly however, patients and clinicians have access to advanced machine learning-based tools for diagnosis, prediction, and recommendation based on large amounts of data, some of it patient-generated. Consequently, just as organizations have had to deal with a "Bring Your Own Device" (BYOD) reality [5] in which employees use their personal devices (phones and tablets) for some aspects of their work, a similar reality of "Bring Your Own Algorithm" (BYOA) is emerging in healthcare with its own challenges and support demands. BYOA is changing patient-clinician interactions and the technologies, skills and workflows related to them.

In this Viewpoint we argue that: (1) BYOA is changing the patient-clinician relationship and the nature of expert work in healthcare, and (2) better patient-clinician-information-interpretation relationships can be facilitated with solutions that integrate technological and organizational perspectives.

**AI is changing the patient-provider-information-interpretation relationships**

Situations in which patients have direct access to algorithmic advice are becoming commonplace [4]. However, many new tools are based on entirely new "black-box" AI-based technologies, whose inner workings are likely not fully understood by patients or clinicians. For example, most patients with Type 1 diabetes now use continuous glucose monitors and insulin pumps to tightly manage their disease. Their clinicians carefully review the data streams from both devices to recommend dosage adjustments. Recently, however, new automated recommender systems to monitor and analyze food intake, insulin doses, physical activity, and other factors influencing glucose levels, and provide data-intensive, AI-based recommendations on how to titrate the regimen are in different stages of FDA approval (e.g., *DreaMed*, *Tidepool Loop*), using "black box" technology - an alluring proposition for a clinical scenario that requires identification of meaningful patterns in complex and voluminous data.

But how these AI-based insights are consumed by the patient and clinician is uncharted territory, with scant population-level evidence to guide its use. Just as Bring Your Own Device can lead to incompatibility between institutional infrastructure and personal tools, with Bring Your Own Algorithm in healthcare, patients and clinicians confront cases where the AI-based advice patients obtain on their own is incompatible with best practice clinical guidelines, the clinician's judgment, or in some cases, with prior models or algorithms used for similar medical cases [2]. Navigating the conflicting recommendations from population-level guidelines and individualized, algorithmic recommendations generated through a combination of advanced medical testing, patient-generated data, and AI-based systems is a challenge for which both clinicians and patients are unprepared. The potential for unproductive contestability [7], where the clinician challenges the machine recommendations that are available to the patient, is concerning because the patient's involvement may transform potentially productive differences in perspective (e.g., clinicians thinking more deeply due to algorithmic advice that differs from their intuition) into personalized conflict that threatens the perceived expertise of the clinician and patient-clinician trust, and may generate uncertainty or worry for the patient. Yet contestability is likely because (a) the machine learning models are fallible and sensitive to bias in training, and (b) patients often lack the broader medical context within which to evaluate the algorithmic advice. As a result, the emerging BYOA reality alters clinicians' role, emphasizing their ability to effectively interact with patients and curate, reconcile and communicate alternative interpretations of the information and recommendation made by algorithmic advice tools. While a wealth of information can help educate patients about their health and medical options, patients often lack the more abstract overarching background that is needed to efficiently interpret the medical information now available to them, leading to misunderstandings or errors that clinicians must correct or reconcile. Troublingly, new tools and misguided interpretation of data can erode patients' trust in clinicians and the medical advice they provide when the AI-based tools offer alternative or conflicting diagnoses, advice, or courses of treatment.

## How we can manage this new reality

As BYOA profoundly alters patient-clinician-information-interpretation relationships, new thinking is required to best harness computing in a clinical interaction context. We see three complementary approaches to potential solutions, bringing together new computing-based tools and organizational practices:

The use of "black-box" tools for diagnoses and recommendation by patients and clinicians begets two undesired outcomes. First, such tools are often not trusted by their clinician users because they do not understand why the tool reached certain diagnoses or recommendations. Clinician distrust may be especially likely in the BYOA situation where the algorithms patients access are unfamiliar to clinicians. Second, increasing patients' direct access to such tools can jeopardize patients' trust in clinicians' judgment and advice [11]. One way to alleviate these concerns involves the use of explainable systems [1], focusing on both user types (patients and clinicians). Much of the research on explainability and interpretability of black-box systems has included visualization of neural networks, analyzing machine learning systems, and training easily-interpretable systems to approximate black box systems. The intended audiences for these approaches are often computer scientists. More work is needed on how explanations should be provided to clinicians (users who do not understand the technology but are experts in the application domain) and patients (users lacking knowledge of technology and application domain). One potential way to make explainable systems more useful is with natural language-based explanation user interfaces, via embodied and non-embodied conversational agents. In previous research [3], we found that there are many complex and interacting human factors that affect non-expert user confidence in a system, including perceptions of the understandability of the explanation, its adequacy, and how intelligent and friendly the system is. The importance of these factors likely differ based on user level of domain expertise, suggesting that different explanations would be effective for patients and physicians. We need to further investigate the effects of different explanatory styles on patients and physicians in BYOA contexts in addition to improving techniques for making black-box algorithms more explainable and interpretable. To align the information patients and clinicians are exposed to while considering the vast differences in their expertise and formal education, new tools should be developed providing patients a simplified version of the explainable systems clinicians use, as well as tools and features that can help users determine the reliability of the algorithms used. Such new tools and features will help enhance patients' and clinicians' trust in the algorithms and understanding of their limitations, mitigate potentially unproductive contestability, and help establish a common ground for patient-clinician interaction and enhanced patient trust in clinicians. To complement the development of patient and clinician-facing explainable systems, new occupations may be needed to serve as curators and communication bridges between patients, medical information and clinicians. Just as new technologies in the past often led to the emergence of new occupational categories and the elimination of others [6], BYOA may demand new work functions whose training and day-to-day operation will integrate medical knowledge, basic understanding of machine learning, communication skills and information and curation savvy. These new healthcare team members will be trained to engage with patients around shared BYOA and explainable systems in ways that are empowering to patients without threatening clinicians. Their inclusion in a patient-focused healthcare environment will be a boon to overburdened and increasingly burned-out clinicians [10] who struggle to cope with growing demands on their time.

A complementary approach treats increased patient interaction with self-diagnosis and advice tools as an opportunity to engage patients in designing future tools. BYOA systems can be a clinical healthcare goal rather than an unplanned outcome of consumer product availability, making the interaction between patients, clinicians, information, and interpretation better managed and more effective. Just as companies benefit from the insights of *lead users* [8] who bring important user perspective and novel ideas to the design of tools companies develop, BYOA tools could benefit from patient-clinician design collaborations, in which the needs, expectations and knowledge gaps of patients will come in close contact with the clinicians, designers and medical informaticists who develop better – and better understood – future tools. In the spirit of user-in-the-loop patient-centered co-design (9), patient-clinician-designer co-design of algorithmic advice tools would focus on the design of a customizable tool whose advice content properties and presentation are adjustable to different personas and user preferences, and levels of computer and visualization literacy. Following such co-design, the adjustment of algorithmic advice tools could ultimately be made by the clinician, the patient, or in consultation between them.  Such patient-in-the-loop design processes, in which patients and clinicians interact around developing BYOA prototypes, could help mitigate misguided or wrong patient self-diagnosis and data interpretation, and the stress and anxiety they can provoke.

**A new era of computing in healthcare**

Computing has a rich history of transforming healthcare: from medical imaging to electronic health records to expert systems, computing has been facilitating major shifts in healthcare practices and tools of the trade. With data-intensive and AI-based computing tools increasingly made available directly to patients, computing is once again transforming healthcare, but this time transforming the medical expert profession and the relationship between patients and their healthcare providers. This transformation poses a number of challenges to clinicians that require new thinking about the emerging patient-clinician-information-interpretation relationships. In this Viewpoint we outline some of the key characteristics of this transformation, and possible ways to address the challenges. We acknowledge that potential solutions may require the development of new tools and roles, which may lead to new challenges, such as the need to integrate new tools into clinicians' workflow. We therefore emphasize the need for a combination of technological and organizational perspectives in scoping and developing such tools and workflows, to ensure that any solution will conform to the Hippocratic Oath principle of "first, do no harm".

**Acknowledgement**

This work was supported by the National Science Foundation awards #1928614 and 1928586.

**About the authors:**

*Oded Nov* is a professor of human-computer interaction at the NYU Tandon School of Engineering.

*Yindalon Aphinyanaphongs* is a physician-scientist, director of the predictive analytics unit and an assistant professor of Healthcare Delivery Science at the NYU Grossman School of Medicine.

*Yvonne W. Lui* is a practicing neuro-radiologist and an associate professor and Associate Chair for AI at the Radiology Department, NYU Grossman School of Medicine.

*Devin Mann* is an associate professor of population health at the Grossman School of Medicine and senior director for Informatics Innovation at NYU Langone Health, as well as a practicing internal medicine physician.

*Maurizio Porfiri* is a professor of mechanical and aerospace engineering, and biomedical engineering at the NYU Tandon School of Engineering.

*Mark Riedl* is an associate professor of computer science in the School of Interactive Computing, Georgia Institute of Technology.

*John-Ross Rizzo* is a physician-scientist and the director of Innovation and Technology for the Rehabilitation Medicine Department; he is an assistant professor of rehabilitation medicine and neurology (NYU Grossman School of Medicine), biomedical, and mechanical engineering (NYU Tandon School of Engineering).

*Batia Wiesenfeld* is a professor of management at the Stern School of Business, NYU.